\documentclass{mn2e}

\usepackage{graphicx}
\usepackage{amsmath}
\usepackage{url}

\title[Inefficient acceleration of electrons around $\Theta^1$ Ori C]
{Inefficient acceleration of electrons in
the shocked wind of the massive star $\Theta^1$ Ori C within the Trapezium cluster}
\author[W. Bednarek]
{
W. Bednarek 
\\ 
University of \L \'od\'z, Faculty of Physics and Applied Informatics,
Department of Astrophysics, ul. Pomorska 149/153, 90-236 \L \'od\'z, Poland,
\\
wlodzimierz.bednarek@uni.lodz.pl
\\}
\begin{document}

\date{Accepted . Received ; in original form }

\pagerange{\pageref{firstpage}--\pageref{lastpage}} \pubyear{2015}

\maketitle

\label{firstpage}
\begin{abstract}
Shocked winds of massive stars in young stellar clusters have been proposed as possible sites in which relativistic particles are accelerated. Electrons, accelerated in such an environment, are expected to efficiently comptonize optical radiation (from massive stars) and the infra-red radiation (re-scattered by the dust within the cluster) producing GeV-TeV gamma-rays. We investigate the time dependent process of acceleration, propagation and radiation of electrons in the stellar wind of the massive star $\Theta^1$ Ori C within the Trapezium cluster. This cluster is located within the nearby Orion Nebula (M 42). We show that the gamma-ray emission expected from the Trapezium cluster is consistent with the present observations of the Orion Molecular Cloud by the Fermi-LAT telescope provided that the efficiency of energy conversion from the stellar wind to relativistic electrons is very low, i.e. $\chi < 10^{-4}$. For such low efficiencies, the gamma-ray emission from electrons accelerated in the stellar wind of $\Theta^1$ Ori C can be only barely observed by the future Cherenkov telescopes, e.g. the Cherenkov Telescope Array (CTA).  
\end{abstract}

\section{Introduction}

Winds of massive stars in young clusters are considered since early 80-ties as likely sites in which cosmic ray particles are able to reach TeV-PeV energies (e.g. V\"olk \& Forman~1982, Cesarsky \& Montmerle~1983, V\"olk \& Forman~1982, Webb, Axford \& Forman 1985). In fact, a few of them, e.g. Cyg OB2 (Aharonian et al.~2005, Abdo et al.~2007, Abeysekara~2021, Cao et al.~2021), Westerlund 1 (Abramowski et al.~2012, Aharonian et al.~2022), Westerlund 2 (Aharonian et al.~2007, H.E.S.S. Collaboration et al.~2011), have been recently observed at TeV $\gamma$-rays. These clusters, with typical age estimated on a few mln years, contain massive stars and their final products of stellar evolution, i.e. Supernova Remnants (SNRs) and/or pulsars and Pulsar Wind Nebulae (PWNe). It is not straightforward to indicate which type of objects, either the winds of massive stars (isolated or cumulative forming super-bubbles), which are still on the Main Sequence, or the remnants of stellar evolution (SNRs, PWNe) play the dominant role in acceleration process of cosmic rays.    
In fact, both groups of these objects are able to supply comparable amount of mechanical energy into the cosmic space (of the order of $\sim$10$^{50}$~erg~s$^{-1}$), ejected in the form 
of the stellar winds or the expanding envelopes of supernovae. However, there are clear differences between proprieties of their potential acceleration sites. The shocks caused by the stellar winds differ from shocks around SNRs which also differ from the shocks structures around PWNe. The winds of massive stars are expected to keep comparable velocities during the whole Main Sequence stage. On the other hand, the shells around the SNRs are expected to significantly decelerate on a time scale of a few thousand years. The structures of the outflows around the PWNe are even much more complicated. The inner parts of the PWNe contain 
relativistic plasma from the pulsar. This part is separated by the relativistic shock from the outer part of the nebula which contains non-relativistic outflow which finally interacts with the surrounding material from the supernova explosion and the interstellar matter.
SNRs are expected to accelerate particles to hundreds of TeV or even PeV energies only at very early stages right after the explosion (see e.g. Marcowith et al. 2018, Inoue et al. 2021, Brose et al. 2022.
These differences might have consequences for the maximum energies reached by the particles and their efficiency of acceleration (see e.g. Brose et al 2020, Celli et al. 2019, Aharonian, Yang \& de O\~na Wilhelmi~2019). The way to distinguish between the roles of these two types of potential acceleration sites is to search for the $\gamma$-ray emission from very young stellar clusters in which even the most massive stars are still observed on the Main Sequence stage. Such clusters are still free from compact objects. The lack of significant $\gamma$-ray emission will indicate that the presence of supernovae provide necessary conditions for the acceleration of particles above TeV energies.  

The evolution of an isolated stellar wind bubble within the star forming region has been studied theoretically already a long time ago (Weaver et al.~1977).  Up to now, a simple, one-zone models have been considered for the radiation processes of accelerated particles in the large scale open cluster
(e.g. Bednarek 2007, Maurin et al.~2016). In these models particles (electrons, hadrons) are injected 
into the dense H II regions of the cluster, with typical size of $\sim$10 pc. These particles are accelerated/re-accelerated on multiple shocks formed by the whole population of massive stars within the cluster. 
In the present work, we consider the high energy radiation produced by electrons accelerated in the wind shock of the massive star which dominates the energy output of the stellar cluster.
  
The interesting candidate for the application of our model is the Trapezium cluster in the Orion Nebula (M 42) which is one of the closest star formation region  with the radius $\sim$4 pc at the distance of 410 pc (Reid et al.~2009).
The Orion Nebula cluster contains a very young cluster of two thousand stars within the distance of 20 light years. The 4 brightest stars within 1.5 light years form so called the Trapezium cluster.  The most luminous star in the Trapezium cluster, $\theta^1$
Ori C, has the luminosity $2\times 10^5$~L$_\odot$ (O6Vp type) (Simon Diaz et al.~2006). Its radiation produces dominant source of ionization of the gas within surrounding Orion Nebula cloud which mass is 
$\sim2.6\times 10^3~M_{\odot}$ (Pabst et al.~2019).  

We investigate the hypothesis in which particles (electrons, hadrons) are accelerated on a strong shock
formed in the stellar wind from the star $\theta^1$ Ori C as a result of its interaction with the surrounding 
gas of Orion Nebula. However, due to the strong radiation field from this massive star, and also infra-red radiation re-processed
by the surrounding gas, electrons should suffer strong energy losses on the Inverse Compton scattering (ICS) process, producing $\gamma$-rays in the GeV-TeV energy range. 
We follow the propagation of electrons in the wind of the massive star taking into account mentioned above their energy losses and by applying the hydrodynamical model developed by Weaver et al.~1977).
From the comparison of the expected $\gamma$-ray emission with the observational constraints by the satellite and ground telescopes, we constrain the efficiency of electron acceleration in the cavity around massive star
$\theta^1$ Ori C. The consequences of these constraints are discussed.

We have chosen the stellar cluster in which a single massive star dominates. However, often many stars produce
comparable wind cavities which can merge forming a super-bubble around the whole cluster. Possible
acceleration of particles within super-bubbles around young clusters has been considered in e.g. Bykov~(2001),
Bykov~(2014), or recently by e.g. Vieu et al.~(2022) and Blasi \& Morlino~(2023).

\begin{table*}
  \begin{tabular}{lllllllll} 
\hline 
\hline 
$\theta^1$ Ori C    &  L$_\star$      &  M$_\star$    & T$_\star$    &  R$_\star$   &  B$\star$    &  L$_{\rm w}$         & $\dot{M}_{\rm w}$  &  v$_{\rm w}$   \\
\hline
unit                &  L$_\odot$      &   $M_\odot$   &      K              &  R$_\odot$    &   G   &  erg/s             & M$_\odot$/yr       &  km/s  \\
value                    &  $2\times 10^5$ &   33          &   $3.9\times 10^4$  &  $10.6\pm 1.5$ & $1100\pm 100$  & $8\times 10^{35}$ &  $4\times 10^{-7}$ &  2500  \\
\hline 
\hline 
\end{tabular}  
\caption{The basic parameters of the massive star $\theta^1$ Ori C: 
Stellar: the luminosity  L$_\star$ (a), the mass M$_\star$ (b), the surface temperature T$_\star$ (a), the radius R$_\star$ (a), 
the surface magnetic field strength (e), the power of the stellar wind L$_{\rm w}$ (c,d), the mass loss rate $\dot{M}_{\rm w}$ and the wind velocity v$_{\rm w}$ (c,d). 
References: (a) Simon Diaz et al. (2006), (b) Balega et al. (2014), (c) Howarth et al (1989), (d) Stahl et al. (1996), (e) Donati et al. (2002).}
\label{tab1}
\end{table*}

\section{A cavity around an early type star within the stellar cluster}

The Trapezium cluster contains four luminous stars known as $\theta^1$~Ori~A,B,C,D. Around $80\%$ of the cluster ionizing radiation is provided by the most massive member of the cluster, $\theta^1$~Ori~C, which is a binary system of O type star and B type star (Balega et al.~2014  ). The star $\theta^1$~Ori~C is immersed in the region of a low density, hot, X-ray emitting gas which forms a bubble expanding with the velocity of $\sim$13~km~s$^{-1}$  (see Fig.~3 in Pabst et al.~2019). The star produces a fast wind 
which parameters are reported in Table~1.
The external pressure of the hot gas balances the pressure of the stellar wind forming a large scale shock at the distance $R_{\rm 1}$ from the star (see Fig.~1 in Weaver et al.~1977). 
This shock can efficiently accelerate particles since the plasma flows through this shock with large velocity. On the other hand, the shock on the border between
the shocked interstellar gas and the ambient interstellar gas, at $R_2$, is slow (the velocity 
of $\sim 13$~km~s$^{-1}$). The slow outer shock is not able to accelerate electrons to large energies since the acceleration efficiency depends on the velocity of the plasma flow through the shock (see Sect.~3).
The small scale shocks in the wind from $\theta^1$~Ori~C can also appear as a result of the interaction with less massive stars 
(e.g. $\theta^1$~Ori~A and D). These two stars have similar masses producing similar winds. They are also located at the similar projection distance from $\theta^1$~Ori~C. If shocks around these stars are able to accelerate electrons to $\sim$TeV energies, then they can lose efficiently energy on the ICS of thermal radiation from the stellar surfaces. Moreover, the star $\theta^1$ Ori C is itself a compact binary system of massive stars. The main component (C1) has the parameters reported in Table.~1.
The companion star (C2) is the B type star with the mass $11\pm 5$~M$_\odot$ (Balega et al.~2014). 

The age of the Trapezium cluster is estimated to be $<$1~mln years, with the likely value of $\sim$2$\times 10^5$ yrs (Pabst et al.~2019). Therefore, even super-massive stars are expected to be still on the Main Sequence stage. The Orion Nebula M42 (galactic coordinates: $209^o, -19.27^o$) is one of the closest, massive star forming regions to the Sun. 

\begin{figure*}
\vskip 4.5truecm
\includegraphics{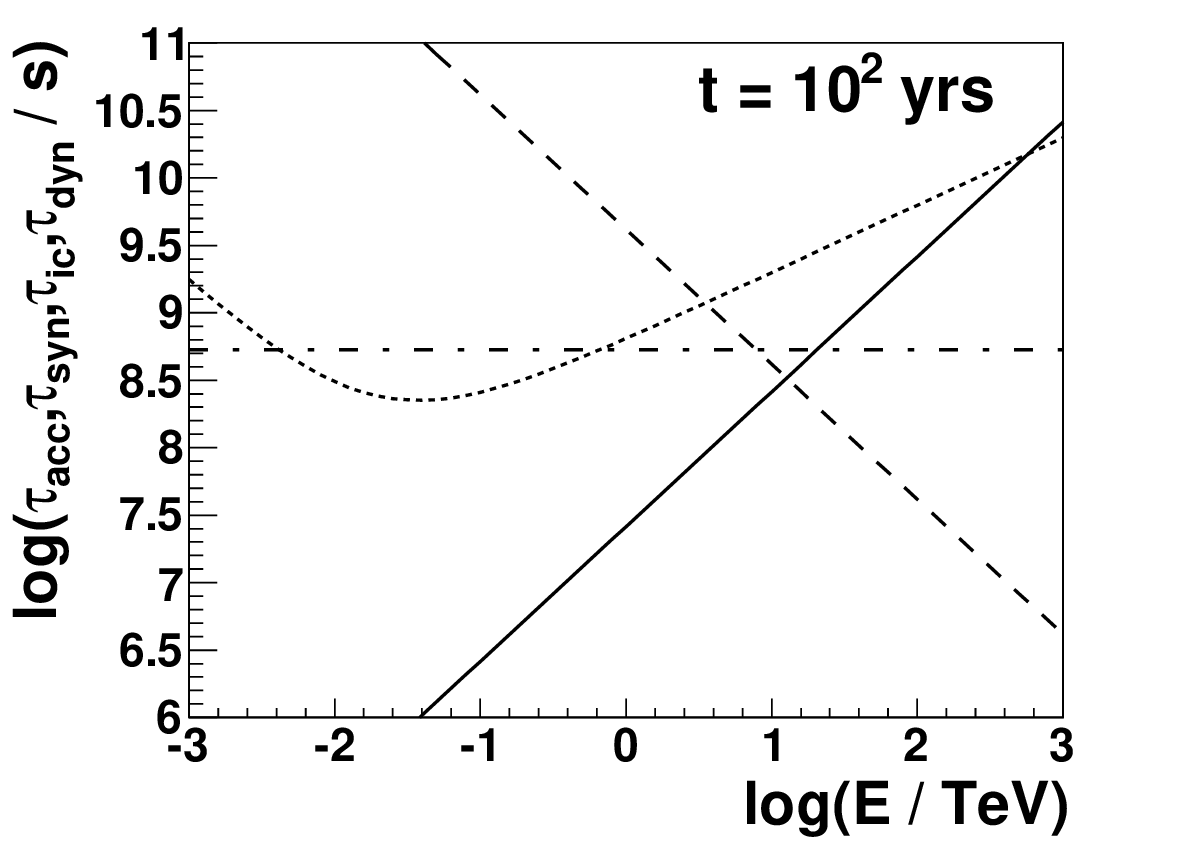}
\includegraphics{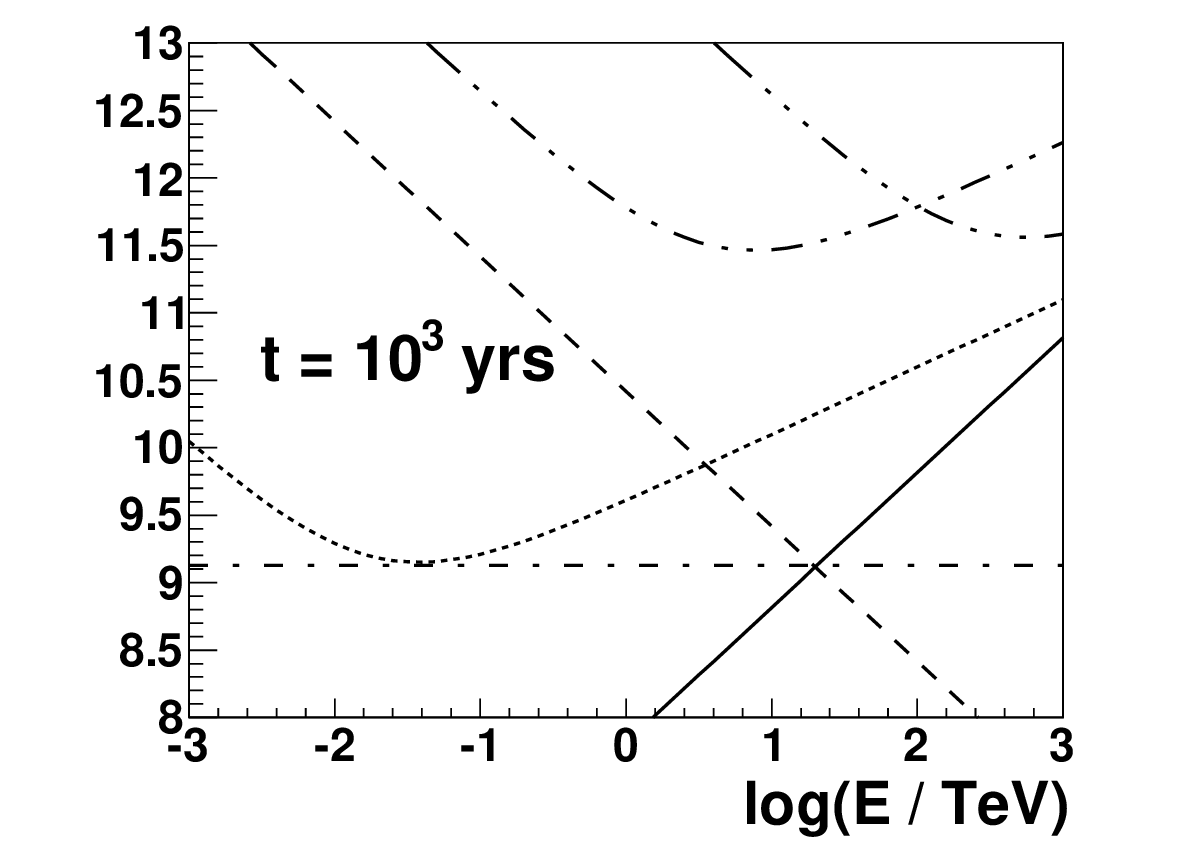}
\includegraphics{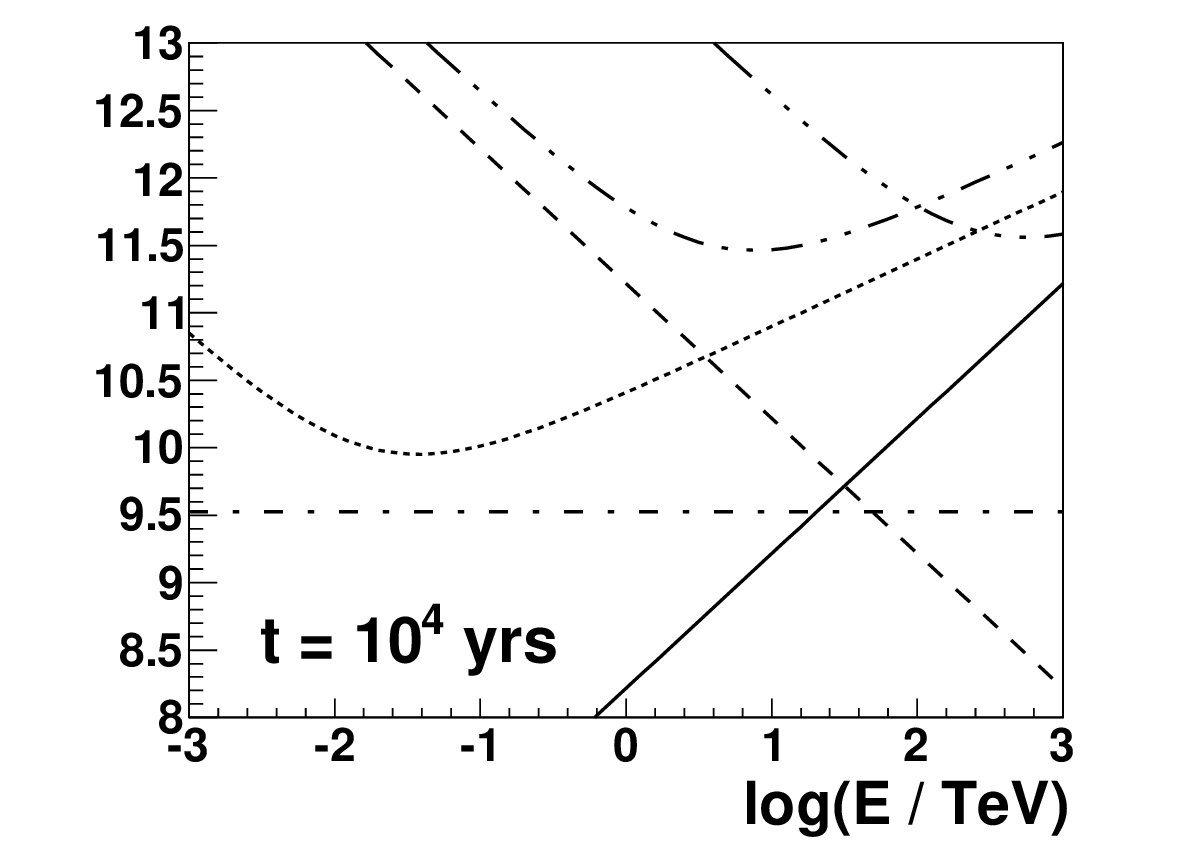}
\includegraphics{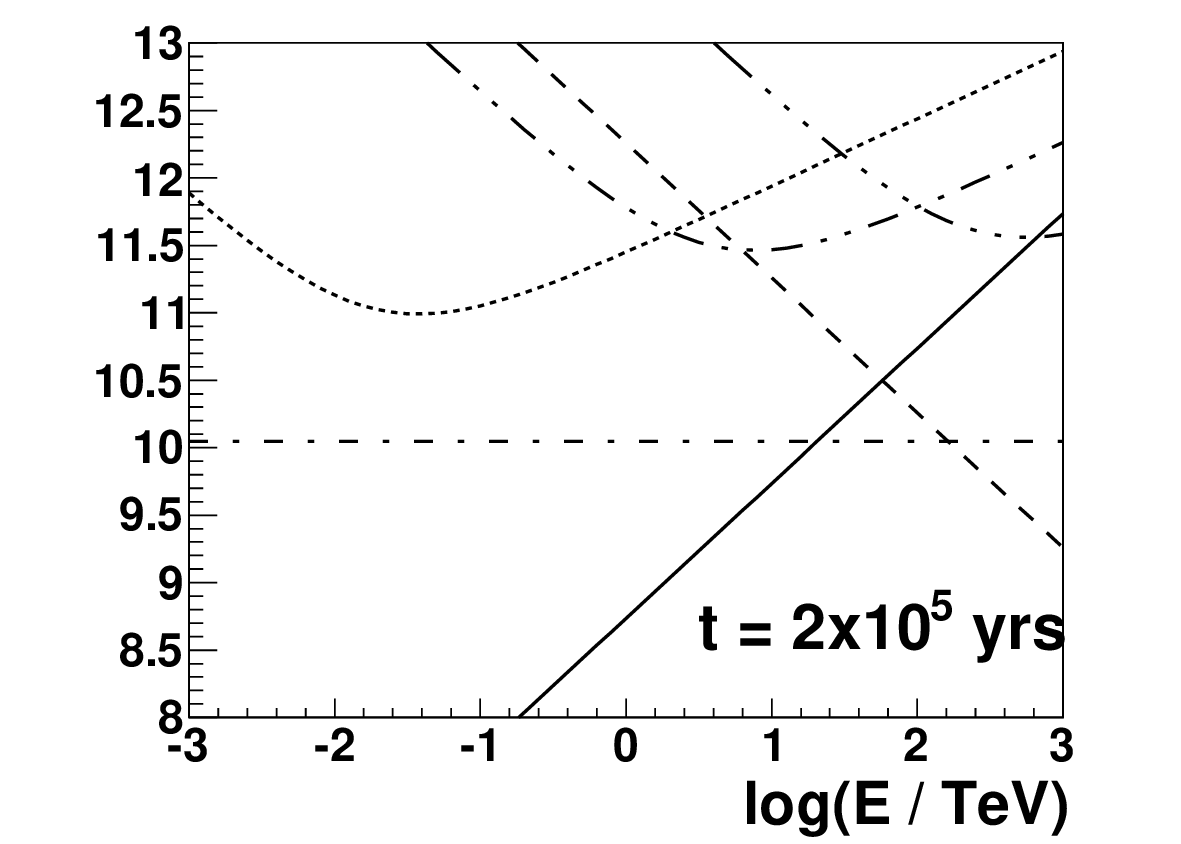}
\caption{The time scale for the energy gains of electrons ($\tau_{\rm acc}$) in the acceleration process 
at the shock in the wind around the massive star $\theta^1$ Ori C is compared with the electron time scales for the energy losses on different process: the advection of electrons from the shock with the velocity of the wind from the star $\theta^1$ Ori C (dot-dashed), the synchrotron energy losses, $\tau_{\rm syn}$ (dashed), and the IC energy losses in the stellar radiation, $\tau_{\rm IC/star}$ (dotted), the infra-red radiation from dust 
$\tau_{\rm IC/inf}$ 
(dot-dot-dashed), and the MBR $\tau_{\rm IC/MBR}$ (dot-dot-dot-dashed) at different moments of the expansion
of the stellar wind nebula counting from its origin: $T = 10^2$ yrs (on the left), $10^3$ yrs (left-middle), $10^4$ yrs (right-middle), and $2\times 10^5$ yrs (right). The surface magnetic field of the star is assumed to be equal to 300 Gs.}
\label{fig1}
\end{figure*}

The powerful wind from the dominant star within the Trapezium cluster, $\theta^1$ Ori C, interacts with the surrounding medium forming a large scale cavity (a bubble) in the Orion Nebula. This cavity is filled with a hot shocked gas with the temperature of the order of 
$T\sim 2\times 10^6$~K  (G\"odel et al.~2008). In the interaction of the fast stellar wind with this surrounding gas a shock structure appears. Detailed structure of the stellar wind cavity, at different moments after its formation, is modelled by Castor et al.~(1975) and Weaver et al.~(1977). This model gives analytic solutions of the hydrodynamic equations for the gas content of the cavity at different stages of its evolution (i.e. as a function of time) and as a function of basic parameters of the stellar wind (wind velocity and the mass loss rate) and the density of surrounding medium. 
Schematic structure of the bubble around a massive star in stellar cluster is shown in Fig.~1 in 
Weaver et al.~(1977). Four regions around the star can be distinguished: (1) the fast stellar wind; 
(2) the shocked stellar wind; (3) the shocked interstellar gas; and (4) the ambient interstellar gas.
Specific parts are separated by the shock in the stellar wind (at $R_1$), the contact surface (at $R_{\rm c}$), and the outer radius of the shocked interstellar gas (at $R_2$). We assume that electrons are accelerated at the strong shock at the distance $R_1$. They are advected to the outer part of the bubble with the velocity of the shocked stellar wind described by Eq.~11 (Weaver et al.~1977). The shocked stellar wind reaches the boundary between different media at the distance from the star $R_{\rm c}$ with the velocity $v(R_{\rm c}) = 3R_{\rm c}/(5t)$, where $t$ is the time after formation of the star. 
For the parameters of $\theta^1$ Ori C, this velocity is close to $\sim$10~km~s$^{-1}$.
For the parameters of $\theta^1$ Ori C, we derive the formulae for above mentioned characteristic distances (see Weaver et al.~1977) in order to model the processes of propagation and radiation of electrons accelerated in the region of the strong shock in the wind region.  

Note however the shortcomings of the Weaver et al. model. For example, it  ignores the effects of the photo-ionization on the dynamics of the wind bubble. The numerical simulations (e.g. Dwarkardas 2022), and also analytic calculations (e.g. Chevalier 1997),
have shown that due to hydro-dynamic instabilities the spherical symmetry of the wind can be destroyed. We do not take these effects into account in our simplified model for the wind bubble dynamics which is considered to be spherically symmetric.
 
Electrons, accelerated at the early time in the bubble (i.e for $t < 100$ yrs), are advected with the shocked stellar wind. Electrons with energies above a few GeV have enough time to cool in the magnetic field and the radiation field of the massive star and the re-processed infra-red radiation fields (see Fig.~1 on the left, for the electron cooling time scales and their advection time scale at the example age of the wind bubble equal to 100 yrs).
These electrons cool already within the stellar wind region below the distance $R_{\rm c}$ (calculated for the age of the wind bubble). Therefore, 
early accelerated electrons are not transported into the dense, shocked interstellar gas.
They are confined close to the wind shock region.     
The soft radiation field within the stellar wind cavity comes mainly from the massive star (optical), from the surrounding cloud (infra-red), and also the Microwave Background Radiation (MBR). The density of the optical radiation is assumed to drop with the distance squared from the star 
and the density of the infrared radiation is kept to be constant within the wind bubble.
The infra-red emission by dust, in the swept up shell around the wind bubble, is $6\times 10^4$~L$_\odot$. The temperature of the dust is $\sim$200 K (Pabst et al.~2019). The dust is heated by the optical emission from $\theta^1$ Ori C. These optical photons propagate freely through the ionized medium of the hot bubble. We calculate the average density of this infra-red photons assuming the dimension of the swept up dust layer equal to $\sim$2~pc.
The wind bubble with such radius around $\theta^1$ Ori C has been measured by Pabst et al.~2019). This value is consistent with the theoretical solution for the wind structure around massive stars (Weaver et al.~1977).

\section{High energy electrons in the stellar wind cavity}

It is assumed that electrons are accelerated at a constant rate (independent on time, 
i.e. the amount of energy transferred from the shock to the relativistic electrons is time independent) at the stellar wind shock from the moment of the formation of the cavity up to the observation time, i.e. after 
$2\times 10^5$ yrs. Electrons are advected from the shock region with the stellar wind. 
During their propagation in the stellar wind,
they lose energy mainly on the IC and the synchrotron processes. Energy losses of electrons depend on time due to the time dependent conditions within the cavity (i.e. the radiation field and the strength of the magnetic field).
We take into account the effects of electron energy losses in the varying medium applying the step time method. 

Let us define the structure of the magnetic field around the massive star which is needed to determine the acceleration process of electrons at the stellar wind shock and their synchrotron energy losses. The inner structure of the magnetic field of the star $\theta^1$ Ori C is assumed to be of the dipole type with an intensity at the surface reported in Table 1 and the magnetic axis inclined at the angle of 
$42^\circ\pm 6^\circ$  with respect to the rotation axis (Donati et al. 2002), i.e. 
$B(r) = B_\star/r^3$, where  $r = R/R_\star$ and $B_\star = 10^3B_3$~G.
The density of the stellar wind at some distance, $R$, from the star is, 
$\rho = \dot{M}_{\rm w}/4\pi R^2v_{\rm w}m_{\rm p}\approx 1.8\times 10^{-14}/r^2$~g~cm$^{-3}$. 
Its energy density depends on the distance from the star as
$U_{\rm w} = \rho v_{\rm w}^2/2\approx 550/r^2$~erg~cm$^{-3}$. This energy density of the stellar wind starts to dominate over the energy density of the magnetic field, $\rho_{\rm B} = B(r)^2/(8\pi)\approx 4\times 10^4B_3^2/r^6$~erg~cm$^{-3}$, at distances larger than the Alfven radius which is equal to 
$r_{\rm A}\approx 2.9B_3^{1/2}$ for the parameters of $\theta^1$ Ori C.
We conclude that the magnetic field of the star has the dipole structure only below the Alfven radius, 
$r_{\rm A}$. At distances above $r_{\rm A}$, the magnetic field structure in the wind is well described by the dominant radial component, $B(r) = (B_\star/(r_{\rm A})(r_{\rm r/t}/r)^2)$ where $r_{\rm r/t} = v_{\rm w}/v_{\rm rot}$ is the radius where the radial magnetic field turns into toroidal.
$v_{\rm rot}$ is the rotational velocity of $\theta^1$ Ori C equal to  53 km/s (Howarth et al. 1997). 
For $\theta^1$ Ori C, we obtain $r_{\rm r/t}\sim 50$.
At large distances from the star, the magnetic field becomes toroidal due to the rotation of the star, $B(r)\approx B_\star/(r_{\rm A}r_{\rm r/t}r)$. Then, the magnetic field strength at large distance from the star $\theta^1$ Ori C (i.e at the radius of the shock $r_{\rm sh}\gg r_{\rm r/t}$) obtains toroidal structure. It can be expressed by $B_{\rm sh}\approx 4\times B(r_{\rm sh})\approx 2\times 10^{-5}B_3/R_{18}$~G for the toroidal magnetic field, where the distance from the star is  renormalized according to $R = 10^{18}R_{18}$~cm.  

\begin{figure*}
\vskip 5.truecm
\includegraphics{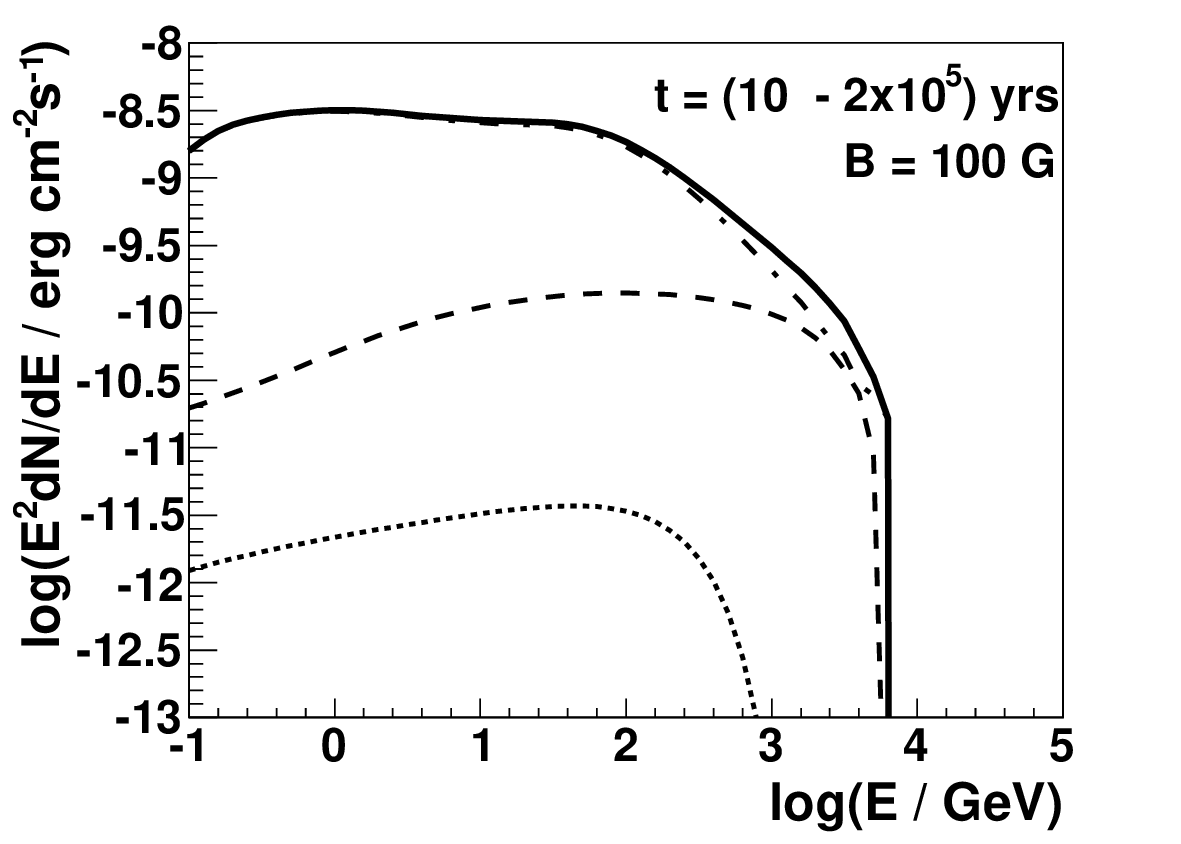}
\includegraphics{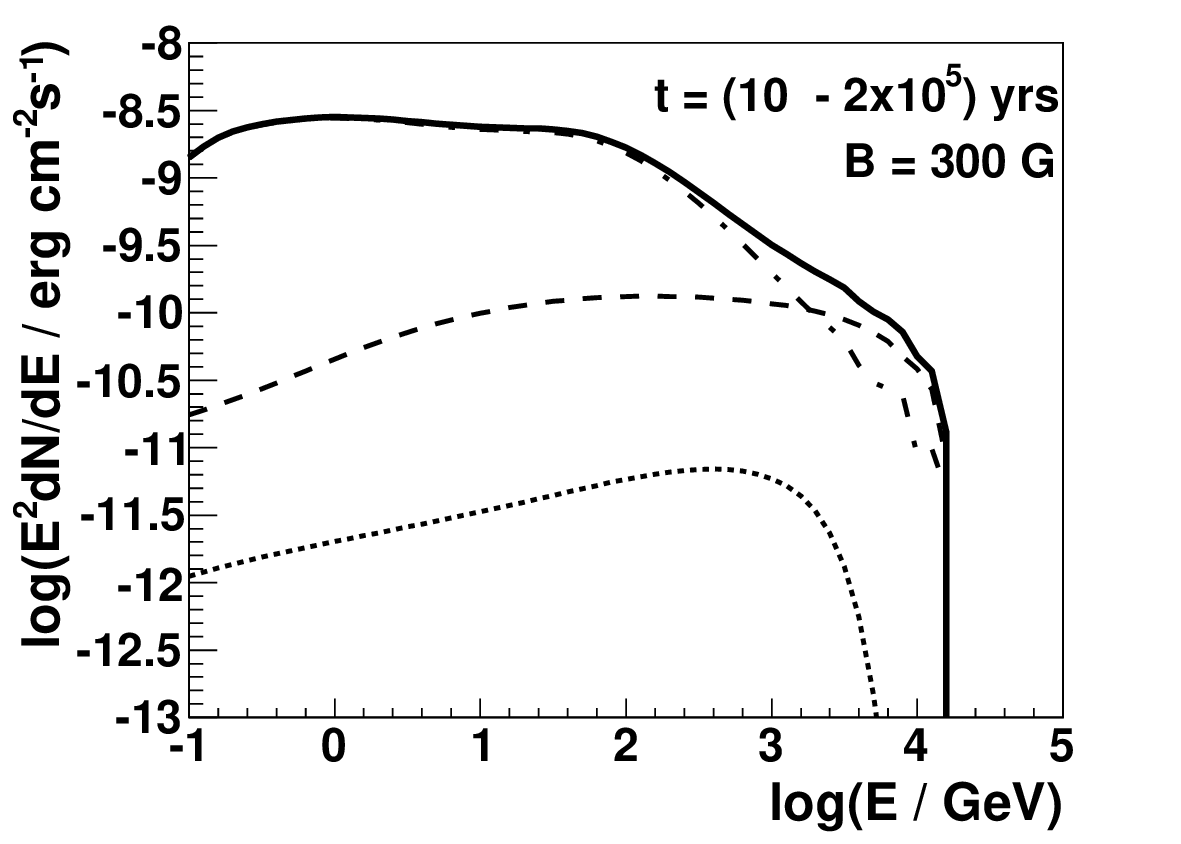}
\includegraphics{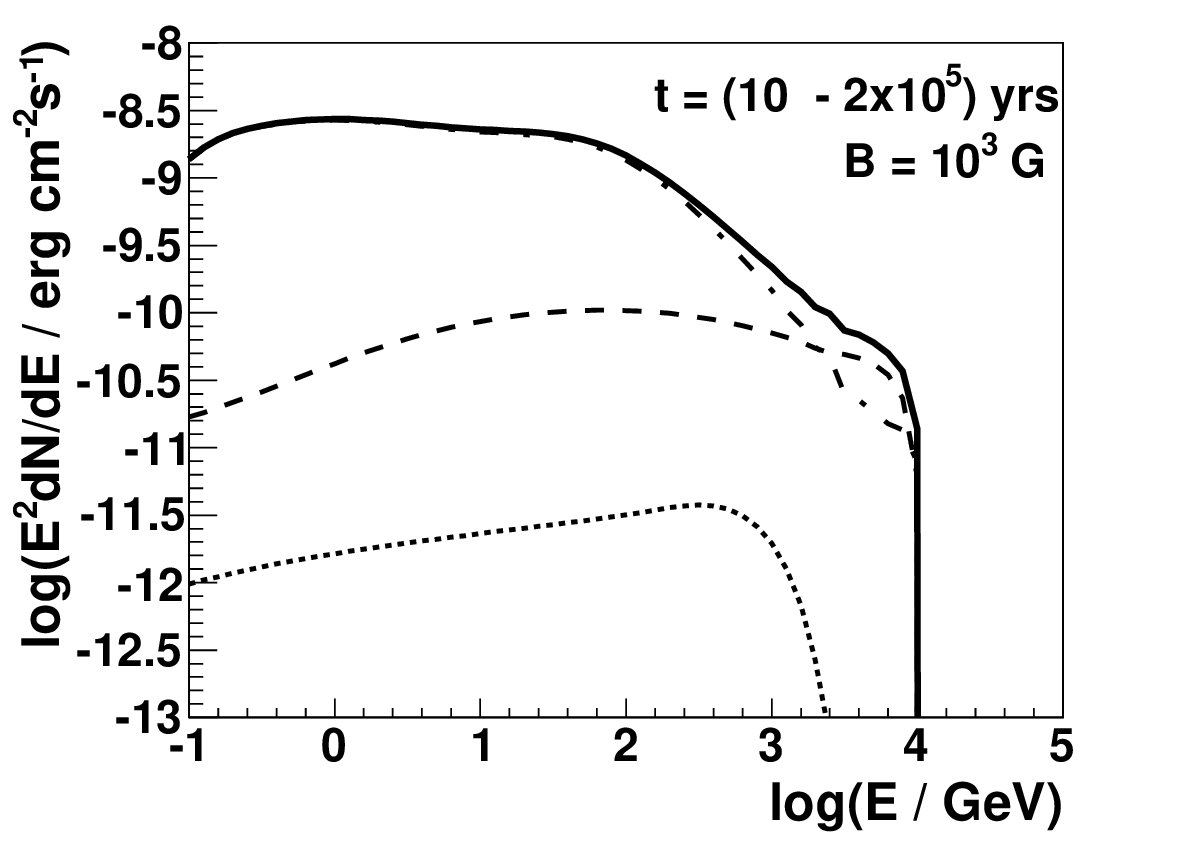}
\caption{Spectral Energy Distribution (SED) of the $\gamma$ ray emission produced by electrons accelerated in the large scale shock around the star $\theta^1$ Ori C. Electrons are injected with the power law spectrum with the spectral index equal to -2, extending from 1~GeV up to $E_{\rm max}$.  $E_{\rm max}$ is determined by the balance between their the acceleration time scale and the dynamical time scale or the synchrotron energy loss time scale (see Fig.~1). In these calculations we accept that $\chi = 0.1$ of the wind power is transferred to relativistic electrons. Electrons are accelerated during the period $t_{\rm min} = 10$~yrs and $t_{\rm max} = 2\times 10^5$~yrs after the formation of the wind cavity. $\gamma$-rays from the comptonization of the stellar radiation (dot-dashed curves), the
infra-red radiation (dashed), and the MBR (dotted). The surface magnetic field of the star 
$\theta^1$ Ori C is $B = 100$~G (figure on the left), 300 G (middle), and $10^3$~G (right).}
\label{fig2}
\end{figure*}

Particles are assumed to be accelerated at the large scale shock in the stellar wind with the power law spectrum up to the maximum energies. In order to determine these maximum energies of electrons,
we compare the time scales for the energy gains of electrons with their time scales for the energy losses on different processes. The energy gain time scale from the acceleration mechanism at the shock can be parametrised by
$\tau_{\rm acc} = R_{\rm L}/\xi/c\approx 10^2E_{\rm TeV}/\xi_{-3}B_{\rm G}$~s, where $R_{\rm L}$ is the Larmor radius of the electrons, $c$ is the velocity of light, 
the energy of the electron is $E_{\rm e} = 1E_{\rm TeV}$~TeV, the magnetic field strength at the shock ($B_{\rm G}$) is expressed in Gauss,
and $\xi = 10^{-3}\xi_{-3}$ is the acceleration coefficient.
The advection time scale of electrons from the shock with the stellar wind is 
$\tau_{\rm adv} = 3R_{\rm sh}/v_{\rm w}\approx 10^{10}R_{18}$~s. 
The synchrotron energy loss time scale of electrons is
$\tau_{\rm syn} = 3m_{\rm e}\gamma_{\rm e}/(4 c \rho_{\rm B}\sigma_{\rm T}\gamma_{\rm e}^2)\approx
141/(B_{\rm G}^2E_{\rm TeV})$~s, where $m_{\rm e}$ is the electron rest mass, $\sigma_{\rm T}$ is the Thomson cross section, 
$\gamma_{\rm e} = E_{\rm e}/m_{\rm e}c^2$ is the Lorentz factor of electrons.  
We have also calculated the energy loss time scale of electrons on the Inverse Compton Scattering of 
the radiation from the massive star $\theta^1$ Ori C, infrared radiation (reprocessed in the gas) and 
the Microwave Background Radiation (MBR) by using the approximate analytic formula from Moderski et al.~(2005). These radiation fields are defined in Sect.~2.
All those energy gains and losses of electrons, at the large scale shock in the wind of $\theta^1$ Ori C, are compared for different moments after the formation of the wind cavity (see Fig.~1). In these calculations, 
the acceleration coefficient of electrons in the shock is assumed to be equal to 
$\xi = 0.3(v_{\rm w}/c)^2$. We estimate that electrons can reach energies of the order of
several up to a few tens of TeV (Fig.~1). At the early stage of the nebula, the acceleration process of electrons is limited mainly by the synchrotron process but at the latter time by the dynamical time of the shock. 

We assume that electrons are accelerated with the power law spectrum, $dN/(dEdt) = K E^{-\alpha}$ (and the spectral index 
$\alpha = 2$) up to the maximum energies as constrained in Fig.~1. $K$ is the normalization factor of the spectrum to the part of the kinetic power of the stellar wind. 
In fact, the spectra of particles accelerated at the stellar wind termination shocks could be much softer in the case of curved shocks around the Sun's type stars. However, it has been shown (e.g. see Sect.~5 in  Webb et al.~1985) that this is not the case for the shocks with the parameters typical for the OB type wind shocks.
For the modelling of the $\gamma$-ray emission, we assume that $\chi = 0.1$ of the wind energy is transferred to relativistic electrons with energies above $E_{\rm e}^{\rm min} = 1$~GeV ($L_{\rm e} = 0.1L_{\rm w}\approx 8\times 10^{34}$~erg~s$^{-1}$). 
Electrons are confined within the advection time at the shock region, losing a part of their energies
on the synchrotron and the IC processes. The electrons in the most energetic part of their spectrum, i.e. above a few TeV (see Fig.~1), lose energy mainly on the synchrotron process. Produced synchrotron radiation extends between $\sim$2~eV up to $\sim$20~keV, for the electrons with energies between $\sim 3-300$~TeV. These electrons can also produce TeV $\gamma$-rays by scattering optical stellar radiation and the infra-red radiation (stellar radiation reprocessed by the gas).
Electrons with energies below $\sim$3 TeV interact mainly in the IC process by scattering optical radiation from the star $\theta^1$ Ori C. They produce $\gamma$-rays with the spectrum extending up to $\sim$TeV energy range (due to the scattering in the Thomson and the Klein-Nishina regimes). 
Our model for the $\gamma$-ray production is applied to the star $\theta^1$ Ori C, since the radiation output and the wind power output of this star dominate the energetics of the Trapezium cluster.

\section{$\gamma$-rays from IC scattering}

\begin{figure}
\vskip 5.5truecm
\includegraphics{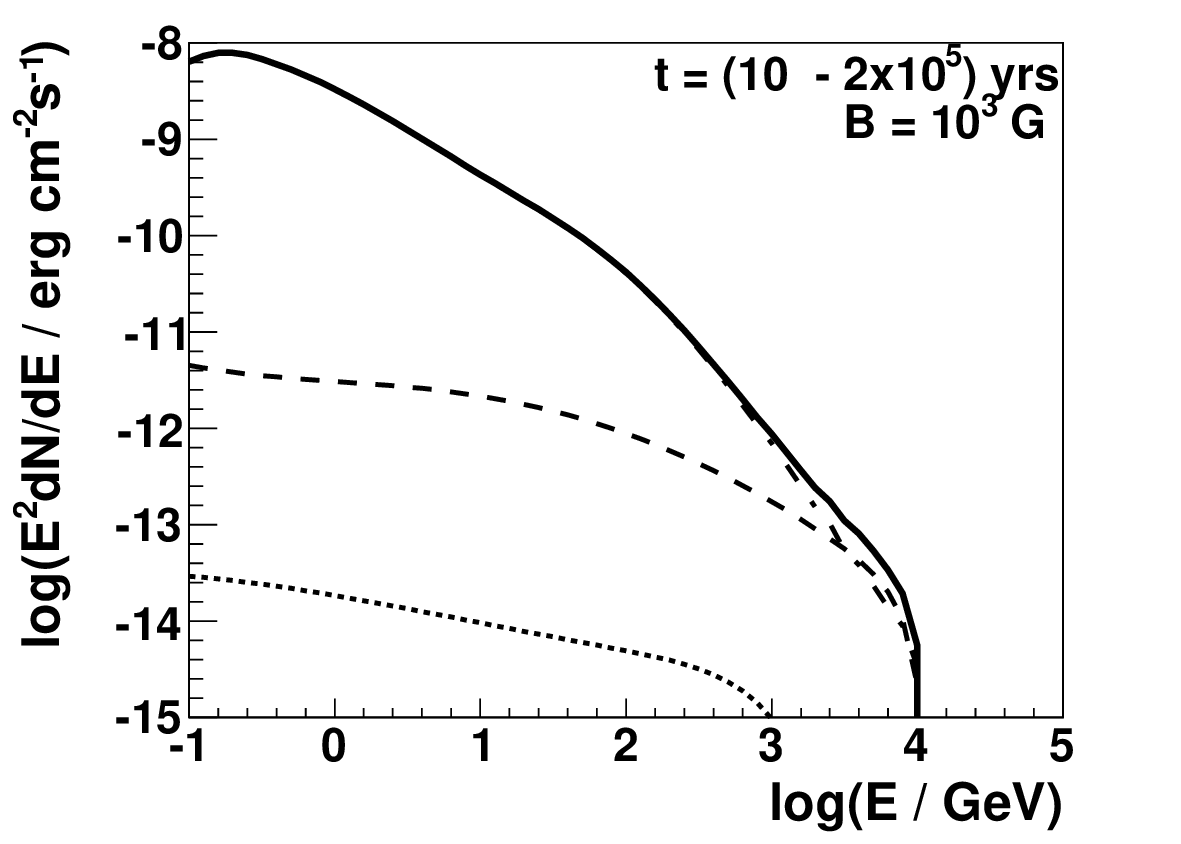}
\caption{As in Fig.~2 but for the spectral index of electrons equal to -3.}
\label{fig3}
\end{figure}

As we have shown above, the energies of electrons injected at the shock in the stellar wind
evolve in time due to their  time dependent energy losses in varying magnetic field and the radiation field within the wind cavity. 
The calculation of the high energy radiation from electrons, which were accelerated in the past, have been done numerically since the physical conditions for these electrons (and so their injected spectrum but not their injected power) depend on time.  
It is assumed that the bubble has the age of $t_{\rm max} = 2\times 10^5$~yrs.
We follow the evolution of the wind bubble from the moment of its formation up to, $t_{\rm max}$, with the step time method in the log scale with the step $\Delta$ log(t). At the  time 't' in the past (counted from the formation of the wind bubble, with the step in time $\Delta~\log(t) = 0.001$ (yr)), we inject relativistic electrons with
the power law spectrum up to the maximum energies which are determined from the comparison of their acceleration time scale with their energy loss time scales or their advection time scale. The spectrum of electrons at the time, t,  is fragmented onto a sequence of bunches with the size in energy  $\Delta \log(E_{\rm (n)}) = \Delta \log(E_{\rm (n-1)}$) + 0.025 (TeV), starting from $\Delta \log(E_{\rm 1})$ = -3, up to the maximum energies allowed by their energy losses.  
The number of electrons in the bunch with energy, E, is calculated from $\Delta N = (dN/dEdt) E (\Delta \log (E)) t (\Delta \log (t)) (\ln 10)^2$.
Different radiation processes such as, the synchrotron and the IC processes in the stellar radiation, infra-red radiation, and also the Microwave Background Radiation (at late periods after formation of the wind bubble), are taken into account.         
The fate of these specific bunches of electrons are followed from the time 't' up to '$t_{\rm max}$', including their energy losses and the location within the bubble at different moments 't'. 
The evolution of the energy of electrons in the bunch is followed in time in linear scale with the time step, 
$\Delta t = 10^{-4}\tau_{\rm loss}$, where $\tau_{\rm loss}$ is the energy dependent time scale for the electron energy losses on considered radiation processes 
(i.e. $1/\tau_{\rm loss} = 1/\tau_{\rm IC} + 1/\tau_{\rm syn}$). In this way, we arrive with a specific electron bunch to the present moment $t_{\rm max}$. Gamma-ray emission is calculated as the sum of emission from all the bunches of electrons which arrived to the time, $t_{\rm max}$ with energies above 1 GeV . So, we do not calculate the spectrum of electrons at the time $t_{\rm max}$ but only radiation from the whole population of the electron bunches which fate is considered in time from the moment of their acceleration, at the time 't', up to the present moment at '$t_{\rm max}$'. 

We calculate the $\gamma$-ray spectra for a few different values of the surface magnetic field of the massive star (see Fig.~2).  
The $\gamma$-ray spectra are dominated by the IC scattering of the stellar radiation with significant contribution of the IC scattering of the infra-red radiation at the largest energy range. The IC scattering of the MBR plays minor role. The spectra show steepening at around $\sim$100 GeV due to the transition between the IC scattering in the Thomson and the Klein-Nishina regimes and also time dependent cooling. The $\gamma$-ray spectra, produced by the electrons injected with a steeper spectrum, show strong emission at GeV energies but low level of emission at TeV energies (see Fig.~3). 
 
We also investigate the contribution to the observed at present $\gamma$-ray spectrum from the electrons which were accelerated at different ages of the wind cavity. The contribution to the electron spectrum is 
shown on the right of Fig.~4. The $\gamma$-ray spectra, produced by these electrons, are shown on the left of Fig.~4. 
It is clear that electrons accelerated early in the wind bubble lose energy very efficiently.
They are not able to contribute to the presently observed spectrum at TeV energy range. However, electrons, accelerated at the bubble with the age $t = 10^4$~yrs, can still contribute significantly to the GeV energy range (see the upper Fig.~4). On the other hand, $\gamma$-rays with energies above $\sim$100 GeV can only be produced by electrons accelerated within the last half of the lifetime of the stellar wind cavity (see the bottom Fig.~4). Therefore, our estimate should be considered as an upper limit.

\begin{figure*}
\vskip 10.5truecm
\includegraphics{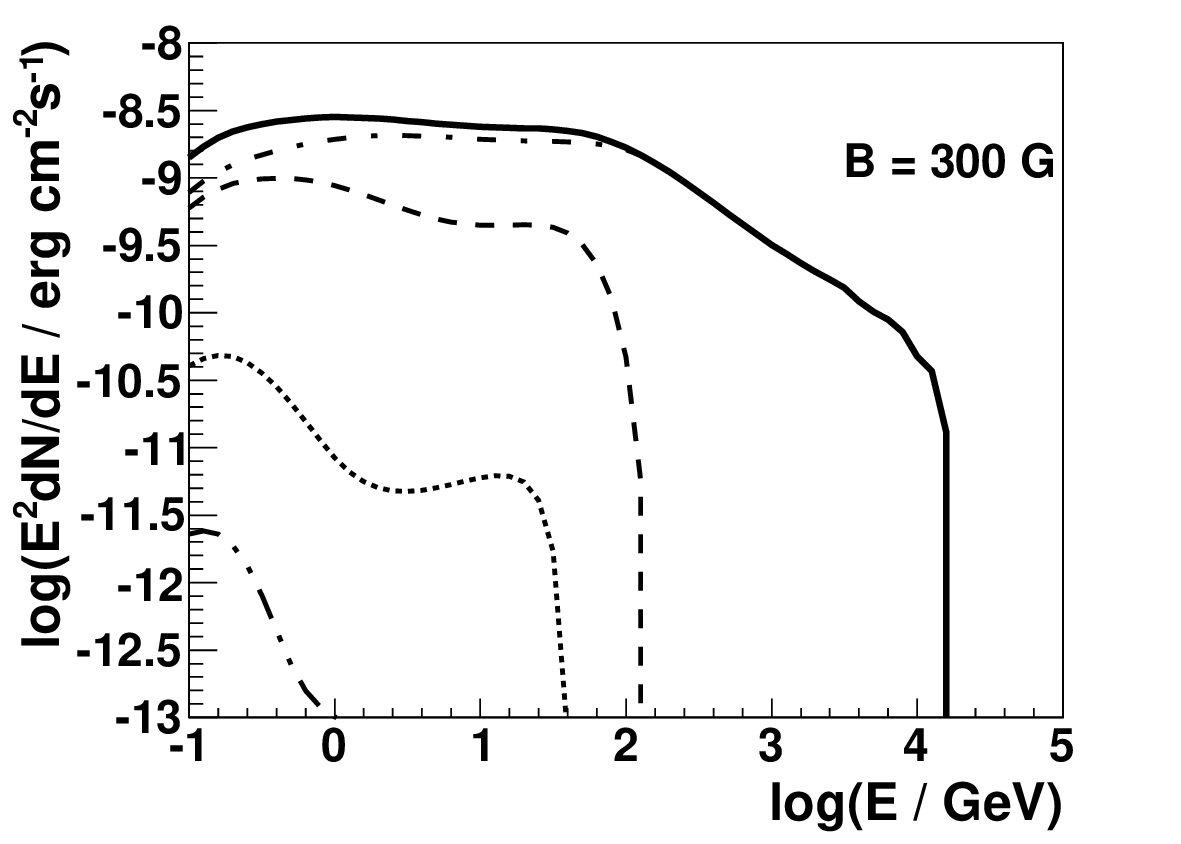}
\includegraphics{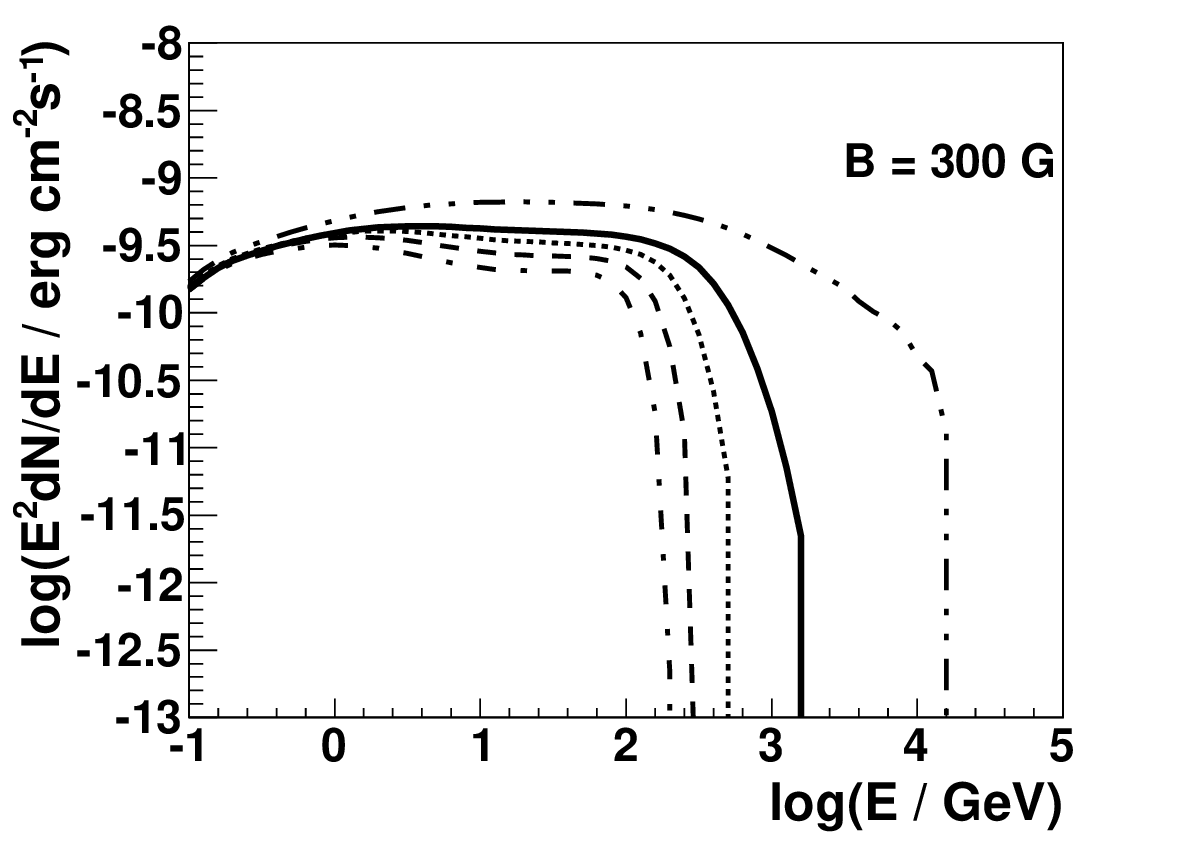}
\includegraphics{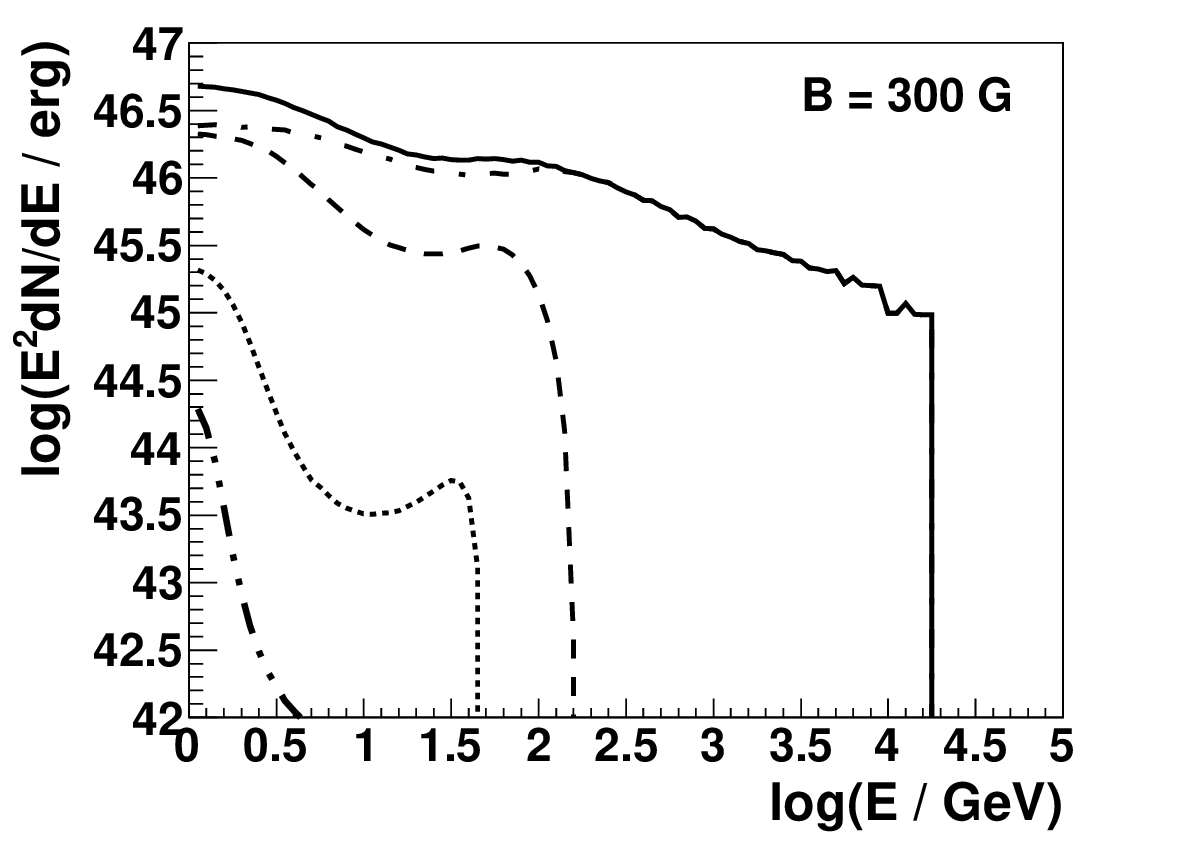}
\includegraphics{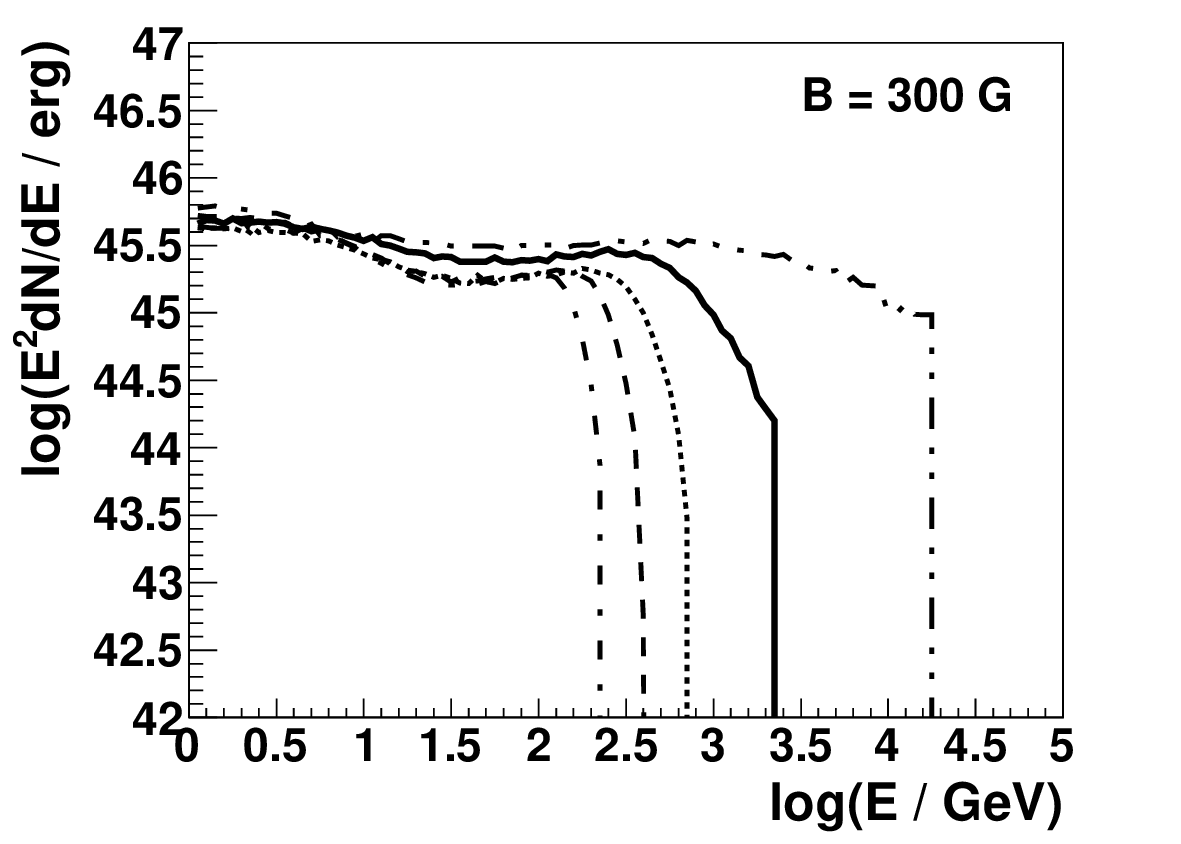}
\caption{Spectral Energy Distribution (SED, figures on the left) of the $\gamma$-ray emission (parameters as in Fig.~2) but for the electrons accelerated at different early periods after formation stellar wind cavity around 
$\theta^1$ Ori C (upper figure):
between $t = 10^5$ yrs and $t = 2\times 10^5$ yrs (dot-dashed curve),  
between $t = 10^4$ yrs and $t = 10^5$ yrs (dashed), 
between $t = 10^3$ yrs and $t = 10^4$ yrs (dotted),
and between $t = 10^2$ yrs and $t = 10^3$ yrs (dot-dot-dashed).   
The $\gamma$-ray emission from the whole period of electron acceleration, between $t_{\rm min} = 10$ yrs and $T_{\rm max} = 2\times 10^5$~yrs is marked by the solid curve.
The emission from electrons accelerated at late periods is shown in the bottom figure,
between $t = 1.85\times 10^5$ yrs and $t = 2\times 10^5$ yrs (dot-dot-dashed curve),  
between $t = 1.65\times 10^5$ yrs and $t = 1.85\times 10^5$ yrs (solid), 
between $t = 1.45\times 10^5$ yrs and $t = 1.65\times 10^5$ yrs (dotted),
between $t = 1.25\times 10^5$ yrs and $t = 1.45\times 10^5$ yrs (dashed),   
and between $t = 10^5$ yrs and $t = 1.25\times 10^5$ yrs (dot-dashed).
The corresponding spectra of electrons, for which these $\gamma$-ray spectra have been calculated, are shown on the right figures.}
\label{fig4}
\end{figure*}

For the stellar wind cavity around the star $\theta^1$ Ori C, with the radius of $R_{\rm neb} = 2$ pc
(Pabst et al.~2019), and the distance of the Trapezium cluster $\sim 410$ pc, the angular size of the TeV $\gamma$-ray emission is equal to $\sim 0.28$ degree. 
This size of the wind cavity is comparable to the diffusion distance of the electrons with energies 1 TeV in the magnetic field at the shock region (estimated on $B = 3\times 10^{-6}$ G, see Sect.~3), assuming the simple Bohm prescription for the diffusion coefficient, which should be considered as an optimistic assumption on the diffusion model of electrons within the wind bubble. However note, that the magnetic field is expected to have toroidal structure at the wind bubble boundary. Therefore, we conclude that in fact electrons accelerated at the stellar wind shock are likely confined within the shocked wind around $\theta^1$ Ori C.

In principle, these $\gamma$-rays might be  absorbed in the radiation from the star $\theta^1$ Ori C.
We estimate the optical depth for the $\gamma$-rays as a function of distance, $D = d\times R_\star$, from the star, on $\tau_{\gamma-\gamma}\approx D\times n_{\rm ph}\times \sigma_{\gamma-\gamma}$, where 
$n_{\rm ph}\approx 1.5\times 10^{15}/d^2$~ph./cm$^3$ is the density of stellar photons as a function of distance from the star for the surface temperature and the radius of the star given in Table~1.  
$\sigma_{\gamma-\gamma}\approx \sigma_{\rm T}/3$ is the cross section for $\gamma-\gamma$ collision. 
Then, $\tau_{\gamma-\gamma}\approx 187/d$. The optical depth for the $\gamma$-rays is below one for distances from the star  $D > 1.4\times 10^{14}$ cm. Therefore, we conclude that $\gamma$-rays, produced at the present location of the shock in the stellar wind region equal to $r_1\approx 10^{18}$~cm (see Eq.~12 in Weaver et al.~1977), do not suffer strong absorption.

\section{Constraints on the acceleration efficiency of electrons}

The Orion Molecular Complex (OMC) is one of the largest, nearby concentration of gas in the solar vicinity. It also contains many interesting stellar objects including the Trapezium cluster. Therefore, it is one of the favourite targets for the $\gamma$-ray observatories. It was detected already by the COS-B (Caraveo et al. 1980) and the EGRET (Digel et al.~1995) Observatories. Recently, OMC is also observed in the GeV $\gamma$-ray energy range by the Fermi-LAT detector (Ackermann et al.~2012, Yang et al.~2014, Neronov et al.~2012, Peng et al.~2019, Baghmanyan et al.~2020). The $\gamma$-ray emission from OMC is extended forming two separate parts OMC A and OMC B. This $\gamma$-ray emission is directionally consistent with the mass concentrations within OMC. Therefore, it is likely due to the interaction of cosmic ray hadrons with the matter of the complex (e.g. see Young et al. 2014).  The upper limit on the TeV $\gamma$-ray emission from the OMC has been 
recently reported by the HAWC Collaboration (Albert et al. 2021). The level of the $\gamma$-ray emission, predicted in terms of our model from the Trapezium cluster (for the electron acceleration efficiency 
$\chi < 10^{-3}$), is in fact clearly below the upper limit on the extended emission from the OMC A (Albert et al.~2021). Recently, the upper limits on the GeV $\gamma$-ray emission from the direction of the Trapezium cluster has been also obtained by Maurin et al.~(2016), based on the analysis of the Fermi-LAT data. Maurin et al.~(2016) considered a hadronic model as responsible for  the $\gamma$-ray emission
from the stellar wind cavity around $\theta^1$ Ori C. In their model, hadrons, accelerated at the stellar wind shock, produce $\gamma$-rays via decay of pions produced in their collisions with the matter of the surrounding gas. They constrained the efficiency of hadron acceleration in the wind of $\theta^1$ Ori C below a few percent.

In Fig.~5, we compare the $\gamma$-ray spectra, expected in terms of our leptonic model from the stellar wind cavity around $\theta^1$ Ori C star, with the upper limits on the Fermi-LAT emission observed towards the direction of the Trapezium cluster (Maurin et al.~2016).
Based on this comparison, we conclude that in order to be consistent with this upper limit, the energy conversion efficiency from the stellar wind to energetic electrons cannot be larger than 
$\chi\sim 10^{-4}$. The efficiency of the electron acceleration is clearly stronger constrained than the efficiency of acceleration of hadrons obtained by Maurin et al.~(2016). This is due to the fact that electrons suffer much stronger energy losses than
hadrons. They are not able to accumulate efficiently within the stellar wind nebula around $\theta^1$ Ori C star. Future observations with the Cherenkov Telescope Array (CTA) will be able only marginally constrain the TeV $\gamma$-ray emission from the wind cavity around $\theta^1$ Ori C star provided that they have leptonic origin (see Fig.~5). Based on the comparison of the Fermi-LAT upper limits with the results of our modelling, we conclude that electrons can not be efficiently accelerated within the wind cavity around the massive star 
$\theta^1$ Ori C.

\begin{figure}
\vskip 5.5truecm
\includegraphics{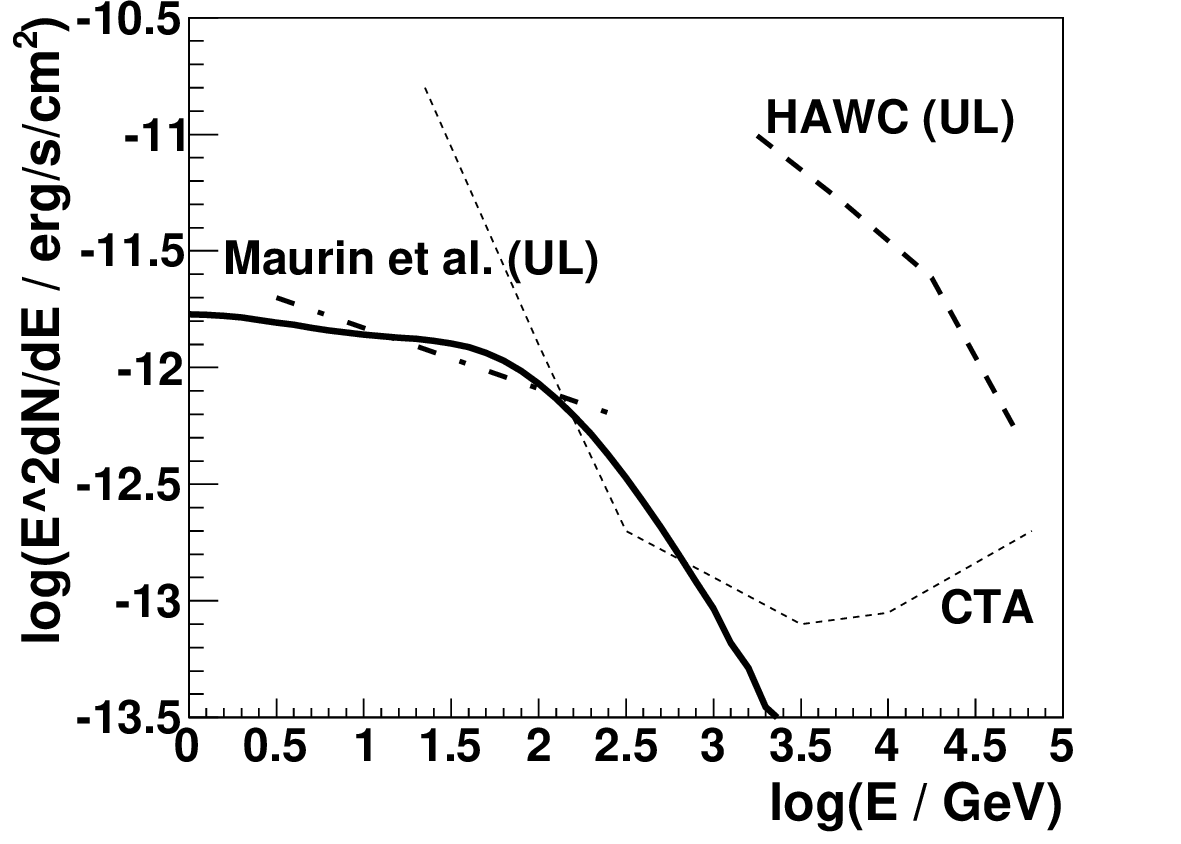}
\caption{Spectral Energy Distribution (SED) of the $\gamma$ ray emission produced by electrons accelerated in the large scale shock around the star $\theta^1$ Ori C (solid curve) is scalled by the electron acceleration efficiency $\chi = 10^{-4}$.  
It is compared with the Fermi-LAT upper limits on the GeV $\gamma$-ray flux from the Trapezium cluster (dot-dashed line, Maurin~et al.~2016) and also the upper limit on the TeV 
$\gamma$-ray emission from the Orion Molecular Cloud obtained by the HAWC~Collaboration (dashed curve, see Table 3 in Albert et al.~2021). We also show the sensitivity of the future Cherenkov Telescope Array (50 hr sensitivity) of the CTA South by the thin dotted curve (Maier et al.~2017).}
\label{fig5}
\end{figure}
\section{The role of shocks around other stars within Trapezium cluster}

The brightest, visible stars in the Trapezium cluster (mentioned above) are in fact close binary systems of massive stars. 
Therefore,  particles (electrons and hadrons), accelerated on the shocks between the winds of these companion stars, could also  produce $\gamma$-ray emission  as discussed by e.g. White  \&  Chen~(1992), Eichler  \&  Usov~(1993), Benaglia \&  Romero~(2003), Torres et al.~(2004), Bednarek~(2005), Reimer et al.~(2006), Pittard  \&  Dougherty~(2006).

In fact, $\theta^1$ Ori C is a binary system of an O type massive star and a B type star with the semi-major axis equal to 
$a = 4.1\times 10^{13}$~cm (for the distance of $D = 415.0\pm 6$~pc), and the eccentricity of 0.6 (Rzaev et al.~2021). The distance between the stars at the periastron is $D = a(1-e)\approx 1.6\times 10^{13}$~cm and at the apoastron is $D = a(1+e)\approx 6.6\times 10^{13}$~cm. The wind parameters of the B type star in this binary system are unfortunately unknown. Therefore, we will apply the typical values for B type stars, i.e. the wind velocity of the order of $v_{\rm w} = 3\times 10^8$~cm~s$^{-1}$, and their mass loss rates of the order of $\dot{M}_{\rm w} = 10^{-9}$~M$_\odot$~s$^{-1}$.
As a result of collisions of these winds, a double shock structure appears within the $\theta^1$ Ori C binary system at the distance from the more massive star equal to, $X_{\rm p,a} = R_{p,a}/(1 + \sqrt{\eta})\approx 0.053R_{p,a}$, where the parameter $\eta$ for the binary system $\theta^1$ Ori C is estimated on, $\dot{M}_{\rm w}^{\rm O}v_{\rm w}^{\rm O}/\dot{M}_{\rm w}^{\rm B}v_{\rm w}^{\rm B}\approx 320$, where $\dot{M}_{\rm w}^{\rm O}$ and $\dot{M}_{\rm w}^{\rm B}$ are the mass loss rates of the O type and B type stars and $v_{\rm w}^{\rm O}$ and $v_{\rm w}^{\rm B}$ are their wind velocities, respectively.
We estimate the solid angle overtaken by the wind of the Be type star as seen from the location of the more massive star in $\theta^1$ Ori C. It is equal to $\Delta\Omega = \pi \theta^2/(4\pi)\sim 7\times 10^{-4}$, where 
$\theta = 1/(1 + \sqrt{\eta})$. Therefore, we conclude that only a very small part of the power of the wind 
from  the more massive star in $\theta^1$ Ori C, of the order of $\Delta\Omega L_{\rm w}\approx 6\times 10^{31}$~erg~s$^{-1}$, can be eventually converted to relativistic electrons in the region of collisions of these winds with the binary system $\theta^1$ Ori C. This is much less than the power available for the electrons accelerated at the large scale shock around the Trapezium cluster considered above. Therefore, we can neglect the role of the acceleration of electrons within the binary system $\theta^1$ Ori C itself.

Other main members of the cluster, i.e. the stars $\theta^1$ Ori A and $\theta^1$ Ori D,
could in principle also form obstacles for the dominant wind from the star $\theta^1$ Ori C.
We assume that B type stars in $\theta^1$ Ori A and $\theta^1$ Ori D
have typical parameters expected for B type stars. The massive star in multiple binary system 
$\theta^1$ Ori A is a B0.5V star with the mass of $15.3$~M$_\odot$. It is at the projected distance of 12.7 mas from $\theta^1$ Ori C. This angular distance corresponds to the physical separation of $7.6\times 10^{16}$~cm, assuming the distance to Trapezium cluster of $\sim$415~pc. The massive companion in $\theta^1$ Ori A star has the radius $4.5$~R$_\odot$, the luminosity $1.6\times 10^4$~L$_\odot$
and the surface temperature $3\times 10^4$~K (Nieva \& Przybilla~2014).
In collisions of the winds from $\theta^1$ Ori C and $\theta^1$ Ori A, the shock structure defined by the parameter $\eta$ equal to $\sim$320 should form. So then, as in the case of the shock around the B type star in the binary system $\theta^1$ Ori C, this shock can extract only a small part of the wind energy from the 
stellar wind of $\theta^1$ Ori C. 
The star $\theta^1$ Ori D has quite similar parameters to $\theta^1$ Ori A, i.e the mass of $18\pm 6$~M$_\odot$ at the projected distance of 13.4 mas from $\theta^1$ Ori C, the radius $5.6$~R$_\odot$, 
the luminosity $2.95\times 10^4$~L$_\odot$, and the surface temperature $3.2\times 10^4$~K (Sim\'on-D\'iaz et al.~2006). Therefore, the effect of the star $\theta^1$ Ori D can be also safely neglected in respect to the radiation processes occurring at the large scale shock around Trapezium cluster.

In the case of strongly magnetized, massive  stars of the $\theta^1$ Ori C type, also particles accelerated in other mechanisms might contribute to the expected $\gamma$-ray emission. For example, particles can be also accelerated in the reconnection region of the stellar magnetic field forced by the strong winds 
from the luminous star (e.g. Usov \& Melrose~1992). Electrons, accelerated in the induced electric field, can produce synchrotron non-thermal radiation (Trigilio et al. 2004, Leto et al. 2017) and also up-scatter dense stellar radiation up to $\sim$TeV $\gamma$-ray energies (e.g. Bednarek 2021). 
We conclude that possible contribution from other processes to the $\gamma$-ray emission from the stellar wind bubble around 
$\theta^1$ Ori C can only farther limit estimated above energy conversion efficiency from the stellar wind to relativistic electrons. 

\section{Conclusion}

We elaborated a model for the acceleration of electrons in the wind cavity around the most massive star in the Trapezium cluster (i.e. $\theta^1$ Ori C) within the Orion Nebula (M 42). We choose a very young cluster (the age $\sim$2$\times 10^5$ yrs) in which even the most massive stars have not exploded yet as a supernovae. In this way, we are able to
investigate the high energy radiation which originates in the winds of massive stars. 
This radiation is not contaminated by the presence of shell type supernova remnants or Pulsar Wind Nebulae and the compact objects (in the binary systems) formed during supernova explosions. It is assumed that electrons are
accelerated at a constant rate during the whole lifetime of the cavity around the most massive star in Trapezium cluster, i.e. $\theta^1$ Ori C. For reasonable acceleration efficiency, electrons can reach TeV energies. However, electrons suffer significant energy losses, due to the synchrotron and the Inverse Compton processes, which strongly modify their spectra observed at present within the stellar wind cavity. Applying the time dependent hydrodynamic model for the evolution of the wind cavity (Weaver et al.~1977), we calculate the $\gamma$-ray emission from electrons at the present moment within the cavity around the star $\theta^1$ Ori C. This $\gamma$-ray emission mainly arises due to the Inverse Compton scattering of the dominant optical radiation from the $\theta^1$ Ori C and the strong infra-red radiation which appears as a result of the reprocessing of the stellar optical 
radiation by the gas surrounding the massive star.

We confront this predicted GeV-TeV $\gamma$-ray emission with the available constraints on the $\gamma$-ray emission from this region of the sky in the GeV and TeV energy range. This $\gamma$-ray emission cannot 
overcome the level of the observed GeV $\gamma$-ray emission from the part of the  Giant Molecular Complex
called Orion A (see e.g. Young et al.~2014). This emission is likely due to the interaction of the galactic cosmic rays with the matter of the huge Orion Nebula. Predicted by us $\gamma$-ray emission from electrons cannot also overcome the upper limits on the TeV $\gamma$-ray emission from the Orion Cluster (obtained by HAWC Collaboration, see Albert et al.~2021). Recently, the upper limit on the GeV emission from the direction of the Trapezium cluster has been also reported by Maurin et al.~(2016).
Based on the comparison of the results of our calculations with these Fermi-LAT upper limits, we conclude that the electron acceleration efficiency, i.e. the conversion factor of the wind power of the star $\theta^1$ Ori C to the relativistic electrons, cannot be larger than $\sim$10$^{-4}$. This indicates that electrons cannot be efficiently accelerated in the wind cavity around the massive star $\theta^1$ Ori C. Predicted level of the TeV $\gamma$-ray emission from the wind cavity around $\theta^1$ Ori C (Fig.~5) can stay marginally within the sensitivity limits of the future Cherenkov observatories (e.g. the Cherenkov Telescope Array).

The constraint on the efficiency of electron acceleration in the wind of $\theta^1$ Ori C is clearly lower than the constraint on the acceleration efficiency of hadrons. This constraint (Maurin et al. 2016), on the level of a few percent, is derived based on a similar scenario but in terms of the hadronic model. The difference in the constraints on the efficiency of hadron and electron acceleration is due to the more efficient energy loses of electrons versus hadrons in the environment around the star $\Theta^1$ Ori C.

\section*{Acknowledgments}
I would like to thank the Referee for many valuable comments.
This work is supported by the grant through the Polish National Research Centre No. 2019/33/B/ST9/01904.

\section*{Data Availability}
The simulated data underlying this article will be shared on reasonable request to the corresponding author.


\label{lastpage}

\end{document}